\documentclass[iop]{emulateapj}

\usepackage{graphics,graphicx}
\usepackage{color}
\usepackage{natbib}

\newcommand{\chandra}{\textit{Chandra}}
\def\arcsec{\hbox{$^{\prime\prime}$}}
\def\deg{\hbox{$^\circ$}}
\newcommand{\oo}{\rm [O\,II]$\lambda\lambda$3727,3729/[O\,III]$\lambda$5007}
\newcommand{\ohb}{\rm [O\,III]$\lambda$5007/H$\beta$}
\def\deg{\hbox{$^\circ$}}

\shorttitle{1321+045 - X-ray cluster}
\shortauthors{Kunert-Bajraszewska et al.}

\begin{document}

\title{An X-ray cooling-core Cluster surrounding a low
  power Compact Steep Spectrum Radio source 1321+045}

\author{M. Kunert-Bajraszewska$^{1}$, A. Siemiginowska$^{2}$,
A. Labiano$^{3}$}

\affil{$^{1}$ Toru\'n Centre for Astronomy, Faculty of Physics, Astronomy
and Informatics, NCU,
Grudziacka 5, 87-100 Toru\'n, Poland}
\affil{$^{2}$ Harvard Smithsonian Center for Astrophysics, 60 Garden St,
Cambridge, MA 02138}
\affil{$^{3}$ Centro de Astrobiologia (CSIC-INTA), Carretera de Ajalvir km.
4, 28850 Torrejon de Ardoz, Madrid, Spain}



\label{firstpage}

\begin{abstract}

  We discovered an X-ray cluster in a {\it Chandra} observation
    of the compact steep spectrum (CSS) radio source 1321+045
    ($z=0.263$). CSS sources are thought to be young radio
    objects at the beginning of their evolution and can potentially
    test the cluster heating process.
    1321+045 is a relatively low luminosity source and its morphology
    consists of two radio lobes on the opposite sides of a radio core
    with no evidence for jets or hotspots. The optical emission line
    ratios are consistent with an interstellar medium (ISM) dominated
    by AGN-photoionization with a small contribution from star
    formation, and no contributions from shocks.  Based on these
    ratios, we classify 1321+045 as a low excitation galaxy (LEG) and
    suggest that its radio activity is in a coasting phase. The X-ray
    emission associated with the radio source is detected
    with $36.1 \pm 8.3$ counts, but the origin of this emission is highly
    uncertain.  The current X-ray image of the cluster does not show
    any signatures of a radio source impact on the cluster medium.
    {\it Chandra} detects the cluster emission at $>3\sigma$ level out
    to $\sim$60\arcsec (~240\,kpc).  We obtain the best fit beta model
    parameters of the surface brightness profile of $\beta=0.58\pm0.2$
    and a core radius of 9.4$^{+1.1}_{-0.9}$\,arcsec.  The average
    temperature of the cluster is equal to $\rm kT =
    4.4^{+0.5}_{-0.3}$\,keV, with a temperature and cooling profile
    indicative of a cooling core. We measure the cluster luminosity
    $L_{(0.5-2 \,\rm keV)} = 3 \times 10^{44}$\,erg~${\rm s^{-1}}$ and
    mass $1.5\times 10^{14}\,M_{\odot}$.  
\end{abstract}


\section{Introduction}
 
Many X-ray clusters are found around radio galaxies with
large-scale radio structures which in majority are classified as
FR\,Is \citep{fr} with a smaller number of FR\,IIs \citep{owen1997}.
These radio sources are old ($>10^{7} $ years) and their long term
interaction with the cluster environment imprinted a rich variety of
structures into the X-ray morphology, such as bubbles, shock fronts
and ripples \citep{mn07, fabian2012}.  However, little is known about
the nature of the X-ray clusters associated with young compact
radio sources (with radio source sizes $<20$\,kpc), namely the
Gigahertz Peaked Spectrum (GPS) and Compact Steep Spectrum (CSS)
objects. These young (age $<10^{5}$\,years) radio sources are believed
to be at the beginning of their evolution \citep{rea96, fanti95}. If
they reside in clusters the inter-cluster medium (ICM) should
have not been impacted by the radio source and we could observe a
primordial X-ray morphology of the cluster medium.  In addition the
observation of the cluster medium can provide important information
about the physical properties and evolution of the radio source
itself.  However, searches for luminous X-ray clusters associated with
GPS and CSS objects were typically unsuccessful
\citep{siem2003,siem2008}.

The only bright X-ray cluster known to host a CSS source is
3C\,186 discovered by \citet{siem05,siem10}.  It is a well formed
cool core X-ray cluster at high redshift, $z = 1.06$.  The central
cluster galaxy hosts a radio-loud quasar with a powerful FRII-type
small-scale radio morphology indicating the initial phase of its
evolution. While expanding into the cluster medium, the young radio
source can potentially supply the energy required to stabilize the
cluster core against catastrophic cooling.  However, the high
redshift location of this source limits the investigation of the
interactions between the radio source and the ICM.

We discovered an X-ray emission from the galaxy cluster, MaxBCG
J201.08197+04.31863 \citep{koester2007} in our {\it Chandra}
observation of a low power CSS radio source, 1321+045 at $z=0.263$.
Here, we present the analysis of the X-ray cluster emission together
with the analysis of the radio, optical and X-ray properties of the
radio source 1321+045.  This is the second CSS source known to be
associated with the large X-ray cluster. It has a different radio
morphology than the 3C\,186 (FR\,I vs. FR\,II-like) and it is much
less luminous in radio. These two CSS sources probe different radio
source properties but in a similar cluster environment.

Throughout the paper, we assume the cosmology with
${\rm
H_0}$=71${\rm\,km\,s^{-1}\,Mpc^{-1}}$, $\Omega_{M}$=0.27,
$\Omega_{\Lambda}$=0.73.

\begin{table*}
{\scriptsize  
\noindent     
\caption[]{\label{annuli} Best Fit Model Parameters for the X-ray cluster$^a$ }
\vspace{-0.4cm}
\begin{center}
\begin{tabular}{ccccccccccccccc}
  \hline
  R$^b$ (arcsec) & Range (arcsec) & Total Counts$^c$ & Net Counts$^c$ &
kT$^d$ (keV)  & Norm$^e$ & n$_e$ (10$^{-2}$ cm$^{-3}$) & S$^f$ (keV cm$^{-2}$) & $t_{c}^g (10^9$yrs)\\
  \hline
  5.3  & 2.0-8.6   & 815.0$\pm 28.5$ & 761.5$\pm29.5$  & 3.9$_{-0.6}^{+0.7}$
  & 0.0186$\pm0.0012$  & 3.9$\pm0.6$ & 33.9$\pm6.6$ & 1.6$\pm0.3$\\\
  11.9 & 8.6-15.2  & 696.0$\pm 26.4$  & 575.9$\pm28.5$ & 4.1$_{-0.7}^{+0.9}$
  & 0.0043$\pm0.0003$  & 1.9$\pm0.4$ & 57.6$\pm13.8$ & 3.3$\pm0.8$\\
  21.8 & 15.2-28.4 & 924.0$\pm 30.4$ & 606.4$\pm 35.2$ & 5.4$_{-0.5}^{+0.6}$
  & 0.0014$\pm0.0001$ &  1.1$\pm0.1$ & 109.2$\pm12.9$ & 6.7$\pm0.7$  \\
  \hline

\end{tabular}
\end{center} 
\vspace{-0.4cm}
$^a$ listed uncertainties are at 68\% for one interesting parameter.
$^b$ The assumed annuli are circular with the mean radius listed in 
the R column and ranges in Range column; $^c$ A number of  counts   
within the 0.5-7~keV energy range.  $^d$ deprojected temperature; $^e$
Normalization for APEC thermal model defined as Norm = $ { {10^{-14}} 
  \over { 4 \pi [D_A (1+z)]^2} } \int n_e n_H dV$ with the abundance  
table set to \citet{anders}; Note that the Norm values given by  {\tt deproject} are normalized 
to a total volume given by the outermost sphere. $^f$ Entropy; $^g$ cooling time.
}
\label{tab:annuli}
\end{table*}

\section{The {\it Chandra} X-ray Observations and Data Analysis}
\label{sec:data}

The \chandra\ ACIS-S observation of 1321+045 was part of a small
snapshot program targeting seven low radio power CSS sources (in preparation).  
It was performed on 2011-12-14.  The source
was placed at the aim point on the back-illuminated ACIS CCD (S3) and
the observation was made in FAINT mode with 1/8 CCD readout to avoid
pileup. 
We used CIAO 4.4 \citep{fruscione2006}
and CALDB 4.5 in all the data analysis and {\it Sherpa} \citep{freeman2001} for 
modeling and fitting ({\tt cstat} with {\tt simplex} method). We
reprocessed the data using {\tt chandra\_repro} to apply the most
recent instrument calibration.  The script runs {\tt
  acis\_process\_events} which applies the sub-pixel algorithm and
gives the best spatial resolution images. After the standard deadtime
correction of 9.4\% the effective exposure time on the source was
9.5\,ksec. The X-ray centroid is located at $\rm R.A. =
13^h 24^m 19^s.646, \, Decl. = + 4^\circ19\arcmin 07\arcsec.45$
(J2000.0).

\subsection{Image Analysis}
\begin{figure*}
\centering
\includegraphics[width=8.5cm,height=6.5cm]{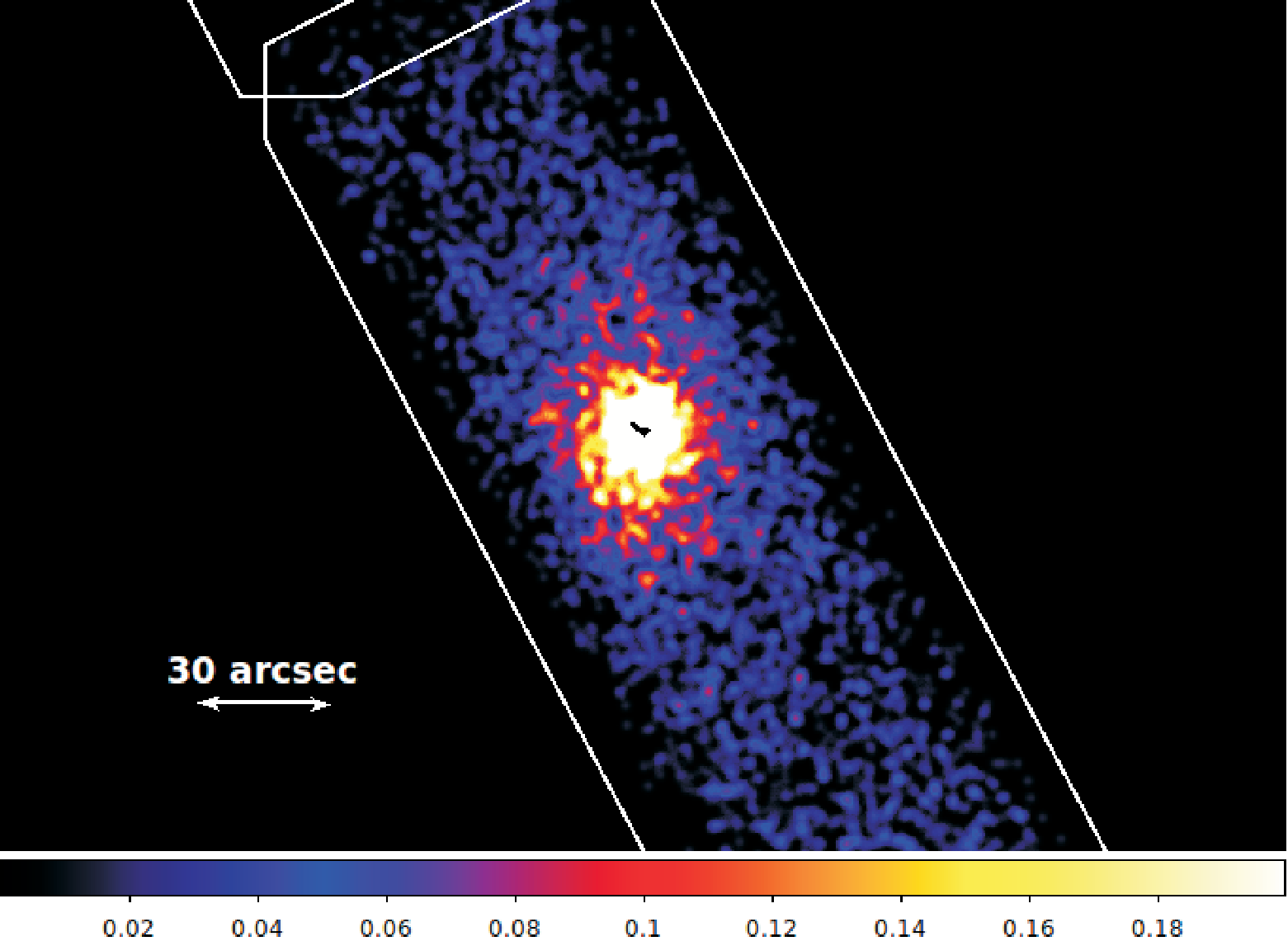}
\includegraphics[width=8.5cm,height=6.5cm]{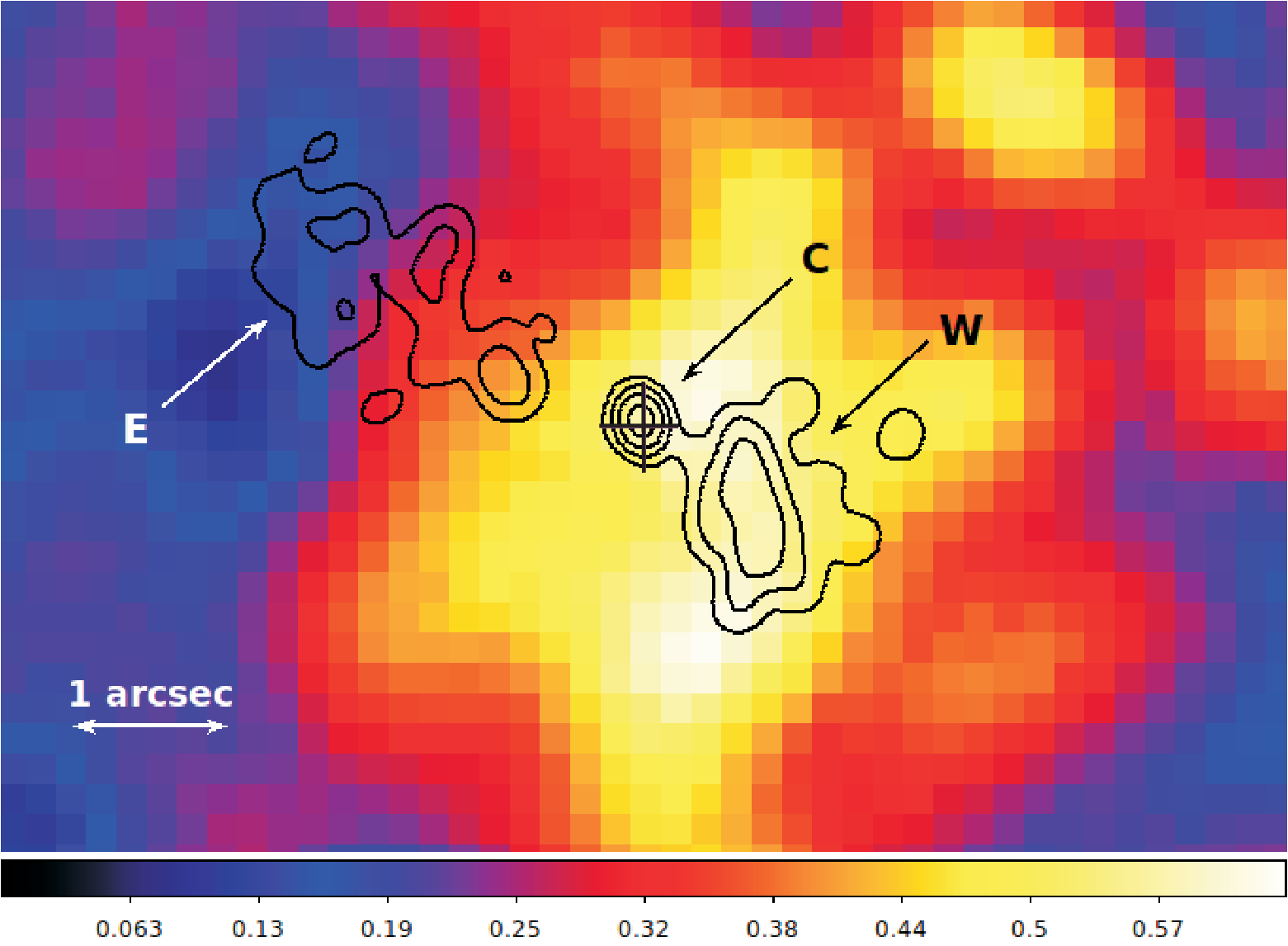}
\caption{{\it Chandra} ACIS-S X-ray image in 0.5-7\,keV energy
  range overlayed with the black radio contours from the MERLIN 1.6\,GHz
image.
The radio contours increase by factor of 2, the first contour level
corresponds to $\approx3\sigma$ and amounts 0.8 mJy/beam.
(Left) The X-ray image smoothed with the Gaussian function
  ($\sigma$=2~arcsec). A 30\arcsec\
  scale bar corresponds to $\sim$121\,kpc.
    The white contours indicate the exposed part of the CCD
    detector.  (Right): The central
  region of the cluster showing the ACIS-S image smoothed with the Gaussian
function ($\sigma$=0.98~arcsec), the radio contours and the optical SDSS
source
  (black cross). The 1\arcsec\ scale bar corresponds to $\sim$4\,kpc.
The radio lobes (E,W) and the core (C) are marked.}
\label{fig:cluster}
\end{figure*}

The X-ray diffuse emission covers a large part of the ACIS-S CCD with
the radio source located in the center (Figure~\ref{fig:cluster}).
The smoothed X-ray image overlayed with radio contours shown in
Fig.~\ref{fig:cluster} seems to suggest a broad enhanced X-ray
emission with two peaks in the vicinity of the core.
Offsets between the radio core and the enhancements are consistent
with the astrometric uncertainty of \chandra\ and we did not attempt
to apply any additional adjustments. The adjustment to the aspect
solution might be possible with a deeper observations in the future if
there are additional point sources detected in this field.

The X-ray cluster emission extends outside the field of view of the
CCD. However, we measured the extent along the CCD using the
surface brightness profiles towards the north and south from the
center. We used {\tt dmextract} and extracted one-dimensional radial
profiles assuming 18 annuli located between 3\arcsec and 100\arcsec
within PA angles of 70\deg-150\deg and 260\deg-335\deg. 
The cluster emission is detected at 3$\sigma$ level out to
about $\sim 60\arcsec$ ($\sim$240\,kpc) from the radio source.
We fit the two profiles in {\it Sherpa} assuming a {\tt beta1d}
model and obtained the core radius of $r_c =
9.4^{+1.1}_{-0.9}$~arcsec and $\beta$-parameter of
0.58$^{+0.03}_{-0.02}$. The extrapolation of the beta1 model into
the circular ($r = 2\arcsec$) region centered on the radio source
shows the excess X-ray emission. We associate this emission with the
radio source.

\subsection{An X-ray emission from the radio core}

We used {\tt specextract} tool to extract the X-ray spectra assuming
1.25\arcsec\ radius circle for the radio source and an annulus with
inner and outer radii equal to 1.5\arcsec\ and 10\arcsec\ respectively
for the local background. 
This region encloses the radio core and a part of the innermost radio
structure.  The X-ray spectrum contains 36.1$\pm8.3$ net counts
(53 total) in the 0.5-7~keV energy range. The cluster emission
contributes to this spectrum and we need to take it into account in
the further modeling of the source spectrum. In the modeling we kept
the absorption parameter at the Galactic value of $\rm N_H = 2.04
\times 10^{20}\,cm^{-2}$ \citep{dickey1990}.

We first fit the background spectrum assuming the APEC model in {\it
Sherpa} (this is the cluster emission within 1.5\arcsec\ and
10\arcsec\ annulus).
We set the metal abundance to 30\% of the Solar values and $z=0.263$
for this model and fit the spectrum containing 1054 counts in
0.5-7\,keV range.  The resulting best fit temperature and
normalization are $\rm kT_b = 4.4^{+0.5}_{-0.3}$~keV and
$8.8\pm0.3\times 10^{-4}$ respectively and represent the average
observed values of the central region of the cluster.  We expect only
about 16$\pm4$ counts from the cluster to contribute to the \chandra\
spectrum of the radio source. These background model parameters
remained unchanged in fitting the spectrum of the radio source
described below.

We assumed an absorbed powerlaw emission model for the radio source
and the APEC model with the fixed parameters to the above best fit values
to account for the cluster emission. The resulting best fit power law
photon index is equal to $\Gamma = 2.35^{+0.39}_{-0.36}$ and the
normalization to $6.2^{+1.4}_{-1.2}\times
10^{-6}$~photons~cm$^{2}$~s$^{-1}$ at 1 keV.  The model unabsorbed
flux of $\rm F_{0.5-2~\rm keV} = 1.4\pm0.3 \times
10^{-14}$~erg~cm$^{-2}$~s$^{-1}$ corresponds to the luminosity $L_{X
  (0.5-2~\rm keV)} \sim 3 \times 10^{42}$~erg~s$^{-1}$ typical for a
low luminosity AGN. The photon index, however, is quite steep and may
indicate a presence of soft thermal emission often observed in low
luminosity AGN and explained as a result of hot ISM of the host galaxy
or a jet emission \citep{hardcastle2009, lamassa2012}. A higher
quality spectrum is needed to understand the origin of this emission.

\subsection{Spectral analysis of the X-ray cluster}

We detected 2882.2$\pm55.4$ net counts within the circular region with
38.2\arcsec\ radius in the \chandra\ 9.7~ksec observation.  The
cluster is bright in X-rays and we attempted to obtain the cluster's
temperature profile by fitting the spectra of three annuli centered on
the radio source. The three annuli span the cluster emission between
2\arcsec\ and 28\arcsec\ and the fourth one between 28\arcsec\ and
35\arcsec\ accounts for the background (see Table 1).

We assume the APEC  model and fit the cluster spectrum in each
annulus.  We account for the cluster 3D volume effects using the {\tt
deproject} model in {\it Sherpa}
\citep[for model details see][]{fabian1981, kriss1983, siem10}.
The best-fit model provides the cluster temperatures,
normalizations, and densities listed in Table~\ref{tab:annuli}.

These numbers suggest that the cluster has a cooling core, although
this result has to be confirmed with better quality data in the
future.  We used the exposure corrected image given by {\tt fluximage}
tool and assumed an elliptical region with 35\arcsec and 60\arcsec
radii to estimate the X-ray luminosity of the cluster to be $L_{(0.5-2
  \,\rm keV)} = 3 \times 10^{44}$\,erg~${\rm s^{-1}}$. The mass of the
cluster enclosed by a sphere with 60\arcsec radius is about $
1.5\times 10^{14}\,M_{\odot}$ assuming the average cluster temperature
of 4.4~keV, and $\beta=0.58$, which is in a broad agreement with the
clusters scaling relations in \cite{eckmiller2011}.

\section{Discussion and conclusions}
\label{sec:discussion}

\subsection{Low Power Radio Source}

The radio source 1321+045 (R.A.~=13$^{\rm h}$24$^{\rm m}$19$^{\rm s}$.7,
Decl.~=~+04$\degr$19$\arcmin$07.2$\arcsec$ (J2000.0))
belongs to a class of young CSS radio sources \citep{fanti95}.  It has
been observed with MERLIN at 1.6\,GHz in 2007 as a part of large
sample of low luminosity compact (LLC) sources \citep{kunert10a}.  The
position of the central component visible in 1.6\,GHz MERLIN image is
well correlated with the position of the optical counterpart
suggesting it is a radio core (C). The two lobes (E and W) are located
on the opposite sides of the core. There is no evidence of jets or
hotspots, however, this needs to be confirmed by observations at
higher frequency.  A total projected length of the source is equal to
$\sim 17\,$kpc and its radio luminosity, $ L_{5{\rm GHz}} \sim
10^{25}$\,W~${\rm Hz^{-1}}$
($< 10^{42}$\,erg~${\rm
  s^{-1}}$), places it in the FR\,I-FR\,II transition region.

Studies of compact radio sources suggest that they exhibit periodic
activity on timescales of $10^{4}-10^{5}$ years \citep{rb97}. On
shorter timescales the radio source is not able to escape from the
host galaxy and starts to recollapse within the ISM \citep{czerny09}.
Our analysis of the whole sample of LLC sources suggest that they can
represent a population of short-lived objects and undergo this phase
of activity many times before they become large scale FR\,I or FR\,II
\citep{kunert10a, kunert10b}. What is more, the evolution of the radio
source and its radio morphology is determined by the properties of the
central engine: strength, accretion mode, excitation level of the
ionized gas, and the ISM.  Optically many of them belong to the class
of low excitation galaxies (LEGs), which are thought to be powered by
the accretion of hot gas \citep{hard07, butti10} and can be
progenitors of large scale LEGs \citep{kunert10b}.  The Ninth SDSS
Data Release \citep{ahn2012} gives the fluxes of the emission lines
visible in spectrum of 1321+045 and the calculated line ratios log \oo
=$1.15$ and log \ohb =$-0.41$ \citep[see][for definitions]{butti10}
are consistent with 1321+045 being a LEG.

We used ITERA \citep{Groves10} to generate emission-line diagnostic
(or BPT, after \citet{baldwin}) diagrams and to compare the
emission-line ratios of 1321+045 with predictions for starburst
\citep{Leith1999,Dopita06,Levesque2010}, dusty
and dust-free AGN \citep{Groves04a}, and shock
\citep{Allen08} models. We found that the emission line ratios of
1312+045 are consistent with AGN-dominated photoionization with
small contributions from star-formation models. None of the shock
models (with and without precursor) reproduce the flux ratios,
suggesting that jet-induced shocks are not present in the ISM of
1321+045 or their contribution to the ionization of the optically
emitting gas is negligible.

\subsection{Interactions between the Radio Source and the ICM}

The presence of distortions and cavities in the X-ray image of a
cluster can signal interactions between the central galaxy (the radio
source) and the ICM \citep{hlavacek2012}. Some clusters have multiple
pairs of cavities filled with low frequency radio emission, probably
caused by the previous radio activity phase.  The high resolution
radio observations show that the reborn radio source can reside inside
the cluster core and multiple cavities provide the evidence for the
previous outbursts \citep{sul12,clarke2009,tremblay2012}.
Figure~\ref{fig:cluster} shows X-ray emission from the 1321+045
cluster in 0.5-7\,keV energy range overlayed with radio contours from
the MERLIN 1.6\,GHz observations.  There is no presence of any ripples
and discontinuities in the current X-ray image of a 1321+045 cluster
but rather uniform emission without indications of an interaction with
the radio jets, or any signature of the previous activity.

Inspection of the VLSS image of 1321+045 at 74\,MHz \citep{cohen}
shows that the low frequency emission extends out to $\sim$120\,kpc.
This low frequency radio emission is uniform
and there is no trace of more extended components which could
indicate the older phase of radio activity.

The synchrotron spectrum of 1321+045 consist of four points and is
steep from 74\,MHz to 5\,GHz with the index $\alpha=0.95$ (defined as
$S\propto\nu^{-\alpha}$). We used a simple model of synchrotron
emission \citep{kunert2009} to reproduce the observed spectrum in
order to find the value of the magnetic field. We took the size of the
radio lobe, 
assumed an equipartition between the particle energy and the magnetic
field energy and a value of the Doppler factor $\delta=1$. We obtained
the best fit value of $B=1.5\times10^{-4}\,{\rm G}$ which gives the
magnetic pressure in each radio lobe to be $\sim 9\times10^{-10}\,{\rm
  dyn\,cm^{-2}}$.  Based on the cluster central density and
temperature we estimate a central thermal pressure of $\sim
5\times10^{-10}\,{\rm dyn\,cm^{-2}}$. Taking into account the
uncertainties in determining the deprojected temperature and density
(only three annuli) and the value of the thermal pressure we conclude
that the cluster environment could limit the growth of the weak radio
source.

Given the value of the magnetic field and the spectrum break
frequency, $\nu_{br}$ we can estimate the synchrotron age of the
source 1321+045. The break frequency indicates the critical point in
which the radio spectrum changes its spectral index and the value of
$\nu_{br}$ depends on the elapsed time since the source formation
\citep{murgia}. In the case of older objects the break frequency is
moved toward lower frequencies.  The synchrotron spectrum of 1321+045
does not show the self-absorption peak or the break frequency in the
range 74\,MHz$-$5\,GHz. We suggest that the well-known electron
self-absorption process modifies the index of the synchrotron
emission of 1321+045 below the 74\,MHz. However, since this assumption
is based on a small number of spectral points we consider two cases in
our calculation of the spectral age of the source: the break frequency
is higher than 5\,GHz or lower than 74\,MHz. This gives us a
synchrotron age of 1321+045 in the order of $\sim10^{5}$ and
$\sim10^{6}$ years, respectively. The $\nu_{br}$ lower than 74\,MHz
indicates that the population of electrons has cooled down to low
energies and the source after a typical for CSS sources lifetime
($10^{4}-10^{5}$ years) started to fade away. As we already suggested
the evolutionary paths of young radio AGNs are probably determined by
the properties of their central engines, namely the HEG/LEG path
\citep{kunert10b}. However, in some objects the surrounding
environment could be also an important factor influencing the
evolution.

\section{Summary}   

1321+045 is the first low power CSS LEG with FR\,I-like radio morphology
discovered to be embedded in an X-ray cluster. The other CSS source,
3C186 \citep{siem05,siem10} has powerful jets and hotspots that are
characteristic of FR\,IIs.  The radio observations of 1321+045 show a
weak radio core and a symmetric diffuse emission from radio lobes
without any compact features. There are no distortions and cavities in
our current X-ray image of the cluster and deeper {\it Chandra}
observations are required to confirm this result.  The optical
analysis rule out the presence of jet-induced shocks in the ISM of
1321+045. We speculate that this low power small scale radio galaxy
did not have enough energy to get out of the host galaxy and it is now
in a coasting phase.

\section{Acknowledgments}

This research has made use of data obtained by the Chandra X-ray
Observatory, and \chandra\ X-ray Center (CXC) in the application
packages CIAO, ChIPS, and Sherpa.  This research is funded in part by
NASA contract NAS8-03060. Partial support for this work was provided
by the \chandra\ grants GO1-12124X.


\begin{thebibliography}{}

\bibitem[Ahn et al.(2012)]{ahn2012} Ahn, C.P., et al., 2012, ApJS, 203, 21 

\bibitem[Allen et al.(2008)]{Allen08} Allen, M.~G., et al., 2008, ApJS, 178, 20.

\bibitem[Anders \& Grevesse (1989)]{anders} Anders E. \& Grevesse N.\
  1989, GeCoA, 53, 197

\bibitem[Baldwin et al.(1981)]{baldwin} Baldwin, J. A., Phillips, M.M.,
Terlevich, R., 1981, PASP, 93, 5

\bibitem[Buttiglione et al.(2010)]{butti10} Buttiglione S., Capetti, A.,
Celotti, A., et al., 2010, A\&A, 509, 6

\bibitem[Clarke et al.(2009)]{clarke2009} Clarke, T.E., Blanton, E.L.;
Sarazin, C.L., et al., 2009, ApJ, 697, 1481

\bibitem[Cohen et al.(2007)]{cohen} Cohen, A.S., Lane, W.M., Cotton, W.D., et al., 2007, AJ, 134, 1245

\bibitem[Czerny et al.(2009)]{czerny09} Czerny B., Siemiginowska, A.,
Janiuk, et al., 2009, ApJ, 698, 840

\bibitem[Dickey 
\& Lockman(1990)]{dickey1990} Dickey, J.~M., \& Lockman, F.~J.\ 1990, ARA\&A, 28, 215 

\bibitem[Dopita et al.(2006)]{Dopita06} Dopita, M.~A., Fischera, J.,
Sutherland, R.~S., et al., 2006, ApJS, 167, 177.

\bibitem[Eckmiller et al.(2011)]{eckmiller2011} Eckmiller, H.~J.,
    Hudson, D.~S., \& Reiprich, T.~H.\ 2011, \aap, 535, A105

\bibitem[Fabian et al.(1981)]{fabian1981} Fabian, A.~C., Hu, 
E.~M., Cowie, L.~L., \& Grindlay, J.\ 1981, \apj, 248, 47 

\bibitem[Fabian (2012)]{fabian2012} Fabian, A.~C., 2012, ARA\&A, 50, 455

\bibitem[Fanaroff \& Riley (1974)]{fr} Fanaroff, B.~L., \& Riley, J.~M.,
1974, MNRAS, 167, 31P

\bibitem[Fanti et al.(1995)]{fanti95} Fanti, C., Fanti, R., Dallacasa, D., et  
al., 1995, A\&A 302, 317

\bibitem[Freeman et al.(2001)]{freeman2001} Freeman, P., Doe, S., 
\& Siemiginowska, A.\ 2001, SPIE, 4477, 76 

\bibitem[Fruscione et al.(2006)]{fruscione2006} Fruscione, A., 
et al.\ 2006, SPIE, 6270,  

\bibitem[Groves \& Allen(2010)]{Groves10} Groves, B.~A \& Allen, M.~G, 2010,
NewA, 15, 614

\bibitem[Groves et al.(2004a)]{Groves04a} Groves, B.~A., Dopita, M.~A.,
Sutherland, R.~S., 2004, ApJS, 153, 9

\bibitem[Hardcastle et al.(2007)]{hard07} Hardcastle, M. J.,
Evans, D. A., Croston, J. H., 2007, MNRAS, 376, 1849

\bibitem[Hardcastle et al.(2009)]{hardcastle2009} Hardcastle, M.~J., 
Evans, D.~A., \& Croston, J.~H.\ 2009, \mnras, 396, 1929 

\bibitem[Hlavacek-Larrondo et al.(2012)]{hlavacek2012} Hlavacek-Larrondo, J., Fabian, A.
C., Sanders, J. S., Taylor, G. B., 2012, MNRAS, 415, 3520

\bibitem[Kunert-Bajraszewska et al.(2009)]{kunert2009} Kunert-Bajraszewska,
M., Siemiginowska, A., Katarzyński, K., Janiuk, A., 2009, ApJ, 705,
1356

\bibitem[Kunert-Bajraszewska et al.(2010a)]{kunert10a} Kunert-Bajraszewska, M.,
Gawronski, M.P., Labiano A., Siemiginowska A., 2010a, MNRAS, 408, 2261

\bibitem[Kunert-Bajraszewska et al.(2010b)]{kunert10b} Kunert-Bajraszewska,
M., \& Labiano, A., 2010b, MNRAS, 408, 2279

\bibitem[Koester et al.(2007)]{koester2007} Koester, B.~P., McKay, 
T.~A., Annis, J., et al.\ 2007, \apj, 660, 239 

\bibitem[Kriss et al.(1983)]{kriss1983} Kriss, G.~A., Cioffi, 
D.~F., \& Canizares, C.~R.\ 1983, \apj, 272, 439 

\bibitem[LaMassa et al.(2012)]{lamassa2012} LaMassa, S.~M., 
Heckman, T.~M., \& Ptak, A.\ 2012, \apj, 758, 82 

\bibitem[Leitherer et al.(1999)]{Leith1999} Leitherer, C., et al., 1999,
ApJS, 123, 3

\bibitem[Levesque et al.(2010)]{Levesque2010} Levesque, E.~M., Kewley, L.~J.,
Larson, K.~L., 2010, AJ, 139, 712

\bibitem[McNamara \& Nulsen (2007)]{mn07} McNamara, B. \& Nulsen \ 2007, ARAA, 45, 117

\bibitem[Murgia et al.(1999)]{murgia} Murgia, M., Fanti, C., Fanti, R., et al., 1999, A\&A, 345,
769

\bibitem[O'Sullivan et al.(2012)]{sul12} O'Sullivan, E., Giacintucci,
S., Babul, A., et al.,2012, MNRAS, 424, 2971

\bibitem[Owen \& Ledlow(1997)]{owen1997} Owen, F.~N., \& Ledlow,
  M.~J.\ 1997, \apjs, 108, 41

\bibitem[Readhead et al.(1996)]{rea96} Readhead, A.~C.~S., Taylor, G.~B., Xu,
W., et~al., 1996, ApJ 460, 612

\bibitem[Reynolds \& Begelman(1997)]{rb97} Reynolds, C. S., \& Begelman,
         M. C., 1997, ApJ, 487, L135

\bibitem[Siemiginowska et al.(2003)]{siem2003} Siemiginowska, 
A., Aldcroft, T.~L., Bechtold, J., et al.\ 2003, PASA, 20, 113 

\bibitem[Siemiginowska et al.(2005)]{siem05} Siemiginowska,
A., Cheung, C.~C., LaMassa, S., 2005, \apj, 632, 110

\bibitem[Siemiginowska et al.(2008)]{siem2008} Siemiginowska, A.,
LaMassa, S., Aldcroft, T. L., et al., 2008, ApJ, 684, 811 

\bibitem[Siemiginowska et al.(2010)]{siem10} Siemiginowska, A., Burke, D.
J., Aldcroft, T. L., et al., 2010, ApJ, 722, 102

\bibitem[Tremblay et al.(2012)]{tremblay2012} Tremblay, G. R., O'Dea, C. P.,
Baum, S. A., et al., 2012, MNRAS, 424, 1026
\end{thebibliography}
\end{document}